\documentclass[notoc]{JHEP3}
\usepackage{amsmath}
\usepackage{epsfig}
\usepackage{amsfonts}
\usepackage{amssymb}
\usepackage{latexsym}
\usepackage{axodraw}
\newcommand{\de}{\partial}

\title{On Anomalies in Orbifold Theories}

\author{Luigi Pilo\\ 
Service de Physique The\'orique, CEA/Saclay, F-91191 Gif-sur-Yvette, France\\
E-mail: \email{pilo@spht.saclay.cea.fr}}
\author{Antonio Riotto\\
Dipartimento di Fisica dell'Universit\`a di Padova and INFN Sezione di Padova
\\ Via Marzolo 8, I-35131 Padova, Italy
\\ 
E-mail: \email{antonio.riotto@pd.infn.it}}
\preprint{Saclay t02/020}
\abstract{We study the issue of gauge
invariance in five-dimensional theories compactified on an orbifold 
$S^1/(\mathbb{Z}_2\times \mathbb{Z}^\prime_2)$ 
in the presence of an external $U(1)$ gauge field. 
From the four-dimensional  point of view the theory contains a tower of Kaluza-Klein Dirac 
fermions  with chiral couplings and it looks anomalous at the quantum level. We show that this
``anomaly'' is cancelled by a topological Chern-Simons term which is 
generated in the effective action when the 
gauge theory is regularized introducing an Pauli-Villars fermion with
an odd mass term. In the presence 
of a   classical background gauge field, the
fermionic current acquires a vacuum expectation 
value, thus  generating 
the suitable Chern-Simons term and a gauge invariant theory.}


\begin{document}
\section{Introduction}

The presence of extra-dimensions is a crucial ingredient in
theories explaining the unification of gravity and gauge forces.
A typical example is string theory where more than three spatial dimensions
are necessary for the consistency of the theory. It has recently
become clear that  extra-dimensions may be very large and could be even 
testable in accelerator experiments \cite{review}. 
Of special interest are  the theories
on orbifold spaces  \cite{orbifold} which are obtained  compactifying the 
extra-dimensions and imposing a discrete symmetry acting on the
higher dimensional coordinates. The four-dimensional 
(4D) low energy effective field theory
coming from an orbifold compactification contains a tower of Kaluza-Klein (KK)
states and may be chiral. The corresponding 4D orbifold 
gauge theory may therefore
be anomalous \cite{adler} -- signaling the breaking of gauge invariance --
unless anomaly cancellation takes place through some non-trivial 
mechanism such as the Green-Schwarz \cite{gs} or the bulk inflow mechanisms
\cite{inflow}. 

The gauge anomaly in five-dimensional (5D) theories compactified 
on   an $S^1/ \mathbb{Z}_2$ orbifold with chiral boundary conditions
for a single bulk fermion with
unit charge under an abelian gauge group $U(1)$
was first  discussed in Ref. \cite{onez}. The anomaly -- 
defined as the five dimensional divergence of the current -- 
lives entirely on the orbifold fixed planes 

\begin{equation}
\label{one}
\partial_M J^M=\frac{1}{2}\left[\delta(y)+\delta(y-\pi R)\right]\,
{\cal Q}(x,y)\, ,
\end{equation}
where $J^M$ is the 5D fermionic current and

\begin{equation}
{\cal Q}(x,y)=\frac{g_5^2}{32\pi^2}\,F_{\mu\nu}(x,y)\widetilde{F}^{\mu\nu}
(x,y)\, ,
\end{equation}
is the 4D chiral anomaly in the external gauge potential $A_M(x,y)$\footnote{
In our notation: $M=\left[(\mu=0,1,2,3),5\right]$ and $y=x^5$
is the fifth coordinate compactified on a circle with radius $R$; $g_5$
is the 5D gauge coupling constant.}.
Therefore four dimensional anomaly cancellation is sufficient to ensure
the consistency of the higher dimensional orbifold theory. 
However,  it was recently claimed that this phenomenon does not persist
 \cite{twoz} in a five-dimensional field theory with a $U(1)$
 gauge field and a charged fermion, compactified on the 
orbifold $S^1/(\mathbb{Z}_2\times \mathbb{Z}^\prime_2)$. Despite the 
fact that the orbifold projections remove both fermionic zero modes, 
gauge anomalies localized at the fixed points were found

\begin{equation}
\label{two}
\partial_M J^M=\frac{1}{4}\left[\delta(y)-\delta(y-\pi R/2)
+\delta(y-\pi R)-\delta(y-3\pi R/2)\right]\,
{\cal Q}(x,y)\, .
\end{equation}
The 4D effective theory is anomaly-free because 
anomalies  cancel after integration over the 
fifth dimension, but gauge invariance is broken, 
spoiling the consistency of the 5D theory. This result 
would be important  for  phenomenologically interesting models
as the one discussed in Ref. \cite{barbieri} whose light spectrum
contains just the zero modes of the 
Standard Model fields with an anomaly-free fermion content.

The goal of this paper is to show that gauge invariance
in models with five-dimensions 
compactified on an orbifold $S^1/(\mathbb{Z}_2\times \mathbb{Z}^\prime_2)$
can be maintained if we admit the presence of an odd mass term for the 
Pauli-Villars regulator.
This procedure leads
to the appearance in the effective action of a Chern-Simons (CS) 
topological term whose gauge variation exactly cancels the anomalous term.

Our results for orbifolds theories are reminiscent of the well-known 
phenomenon present in theories with an odd number of non-compact 
dimensions where the parity-violating part of the vacuum current induced 
by the classical background gauge field implies the presence in the
effective action of a CS topological invariant 
which is odd under parity transformations \cite{redlich}.

The paper is organized as follows. In section 2 we summarize the computation leading to
the anomaly in the theory compactified on  $S^1/(\mathbb{Z}_2\times \mathbb{Z}^\prime_2)$
and with a KK tower of 4D chiral fermions 
with opposite parities under the two $\mathbb{Z}_2$'s. In section
3 we discuss the CS counterterm  suitable to cancel the anomaly and in section
4 we show the CS counterterm is induced by a Pauli-Villars regulator. 
Finally, section 5 contains our conclusions.

\section{The 5D vector current}
Consider a 5D fermion (Dirac) living in $\mathbb{R}^4 \times  
{\cal P}$  coupled with a $U(1)$ external gauge field. The compact 
component of the space is a circle $S^1$ of radius $R$ modulo some  discrete 
group $\mathbb{Z}_2 \times \mathbb{Z}_2^\prime$ (or $\mathbb{Z}_2$)
 which in general 
has fixed points. As a result,
the resulting space in an orbifold. 
The action we consider is 
\begin{equation}
S = \int d^5x \left[ i \bar{\Psi}  \Gamma^{{}_{M}} \de_{{}_{M}} \Psi- g_5 \, \bar{\Psi} \Gamma^{{}_{M}}\Psi  \, A_{{}_{M}}  \right] \quad 
\label{act}
\end{equation}
where 
\begin{equation}
\Gamma^\mu = \begin{pmatrix} 0 &\sigma^\mu  \\
\bar{\sigma}^\mu & 0 \end{pmatrix}, \qquad \Gamma^4 
= -i \begin{pmatrix} -1 & 0 \\
0  & 1 \end{pmatrix} = -i \, \gamma^5,
\quad \sigma^\mu = (1, \vec{\sigma}), \; \; \bar{\sigma}^\mu = 
(1, -\vec{\sigma})  \, .
\end{equation}
The two orbifold projections act on the space-time 
points as 
\begin{equation}
\begin{split}
\mathbb{Z}_2 : \quad & (x^\mu, \, y) \rightarrow (x^\mu, \, y^\prime)  = 
(x^\mu, \, 2\pi R-y) \\ 
\mathbb{Z}_2^\prime  : \quad & (x^\mu, \, y)  \rightarrow (x^\mu, \,
y^\prime) = (x^\mu, \, \pi R-y) \, \qquad y\in [0,2\pi R].
\end{split}
\end{equation}
and $y$ is identified with $y^\prime$. On the 5D spinor the following 
condition is imposed
\begin{equation}
\Psi(x,y^\prime) = \epsilon \, \gamma^5 \, \Psi(x,y) \, , \qquad \epsilon
= \pm 1 \quad .
\label{bc}
\end{equation}
We shall denote by $\epsilon$ and $\epsilon^\prime$ the parity of the field. 
Notice that a 5D mass term $M$ is forbidden unless it has a non-trivial profile in the 
bulk with parities $(-,-)$.
In \cite{twoz} was claimed that the 5D current 
\begin{equation}
J^{{}_M} = \bar{\Psi} \Gamma^{{}_M} \Psi 
\end{equation}
though is classically conserved, has an anomalous divergence at the quantum level.
The argument is based on the result for the covariant anomaly for chiral 
fermions in four 
dimensions. Indeed one can rewrite the action (\ref{act}) as a 
collection of 4D massive Dirac 
fermions by expanding  $\Psi$ in terms of the complete set formed by the 
solutions of free Dirac equation in 5D    
\begin{equation}
\begin{split}
& \Psi(x,y) = \sum_n \left[ \psi_{n{}_R}(x) \, \phi^{{}_R}_n(y) + 
\psi_{n{}_L}(x) \, \phi^{{}_L}_n(y) \right] \quad ; \\
& \psi_{n{}_{L/R}} = P_{{}_{L/R}} \psi_n \quad ;
\end{split}
\end{equation}
where $\psi_n$ is a 4D Dirac fermion of mass $M_n$. The KK modes 
$\phi^{{}_{L/R}}_n$ satisfy
\begin{eqnarray}
\label{h}
&&  \phi^{{}_R}_n \, M_n   - \, \frac{d}{dy} \phi^{{}_L}_n = 0 \quad ;\\
\label{hh}
&& \phi^{{}_L}_n \, M_n + \, \frac{d}{dy} \phi^{{}_R}_n = 0 \quad .
\end{eqnarray}
From Eq. (\ref{bc}) we get the following transformation rules under
the orbifold projections  
\begin{equation}
\phi^{R/L}(-y) = \pm \epsilon \, \phi^{R/L}(y) \,  \qquad
\phi^{R/L}(\pi R -y) = \pm  \epsilon^\prime \, \phi^{R/L}(y)   \quad .
\end{equation}
The action (\ref{act}) can be written, using the orthogonality of the 
KK modes, as 
\begin{equation}
\begin{split}
S = \int d^4x &\left[\sum_n \left(\bar{\psi}_n i \gamma^\mu \de_\mu \psi_n
- M_n \, \bar{\psi}_n  \psi_n \right) \right.\\
&\left. - g_5 \, \sum_{n,m} \left(j^\mu_{{{}_{L}} mn}
\, {\cal A}_{\mu \, mn}^{{}_L} \, 
+ j^\mu_{{{}_{R}} mn} \, {\cal A}_{\mu \, mn}^{{}_R} \, 
-i  \, j_{5 \, mn} \, {\cal A}_{5 \, mn}\right) \right] \quad ;
\end{split}
\label{4dact}
\end{equation}  
where
\begin{equation}
\begin{split}
&{{\cal A}_\mu^{{}_{L/R}}}_{mn} = \int_0^{2 \pi R} dy \;  
\phi^{{}_{L/R}}_n(y) \phi^{{}_{L/R}}_m(y) \, A_\mu(x,y) \quad ;  \\ 
&{{\cal A}_5}_{mn} = \int_0^{2 \pi R} dy \;  
\phi^{{}_{L}}_m(y) \phi^{{}_{R}}_n(y) \, A_5(x,y) \quad ;
\end{split}
\end{equation} 
and 
\begin{equation}
\begin{split}
&j^\mu_{{{}_{L / R}} mn} = \bar{\psi}_m \gamma^\mu 
P_{{}_{L / R}} \, \psi_n \quad ; \\
&j_{5 \, mn} = \bar{\psi}_{{{}_L}m} \, \psi_{{{}_R}n} - \bar{\psi}_{{{}_R}n} \, \psi_{{{}_L}m}
\quad .
\end{split}
\end{equation} 
Let us now reproduce the results in \cite{onez} and \cite{twoz}. After setting $A_5 = 0$ by a 
gauge choice\footnote{Strictly speaking this is not allowed: one is 
using gauge invariance before showing that it is still a good symmetry at the quantum level.},
from the classical equations of motion and the well known result for the anomalous divergence of 
chiral current in 4D one has
\begin{align}
\de_\mu j^\mu_{{{}_L} mn} = i \left[\bar{\psi}_m \, M_m \, 
\, P_{{}_L} \, \psi_n - \bar{\psi}_m \, P_{{}_R} M_n \, \psi_n
\right] - \frac{g_5^2}{32 \pi^2} \left({\cal F}^{{}_L}_{\mu \nu} 
\widetilde{{\cal F}^{{}_L}}^{\mu \nu} \right)_{mn}, \label{q1}\\
\de_\mu j^\mu_{{{}_R} mn} = i \left[\bar{\psi}_m \, M_m \, 
\, P_{{}_R} \, \psi_n - \bar{\psi}_m \, P_{{}_L} M_n \, \psi)_n
\right] + \frac{g_5^2}{32 \pi^2} \left({\cal F}^{{}_R}_{\mu \nu} 
\widetilde{{\cal F}^{{}_R}}^{\mu \nu} \right)_{mn}  .
\label{q2}
\end{align}
On the other hand, the 5D current can written in terms of the 4D currents
\begin{equation}
\begin{split}
&J^\mu(x,y) = \sum_{mn}\left[\phi^{{}_R}_m(y) \phi^{{}_R}_n(y) \, 
j^\mu_{{{}_R} mn}(x) + \phi^{{}_L}_m(y) \phi^{{}_L}_n(y) \,  
j^\mu_{{{}_L} mn}(x) \right]
\quad .\\
& J^5 = -i \sum_{mn} \phi^{{}_L}_m(y) \phi^{{}_R}_n(y) \, 
j_{5 \, mn}(x) \quad .
\end{split}
\label{curr}
\end{equation}
The fifth-dimensional structure of the divergences is recovered noticing that at the classical 
level
\begin{equation}
\de_y J^5  = \sum_{mn} i \left[  M_n \left(\phi^{{}_R}_m \, 
\phi^{{}_R}_n \, - \phi^{{}_L}_m \, \phi^{{}_L}_n \,\right) 
\bar{\psi}_m \, \left(P_{{}_L} - P_{{}_R} \right) \, \psi_n \right] \, .
\label{cls}
\end{equation} 
Combining Eqs.(\ref{q1}), (\ref{q2}), (\ref{curr}) and (\ref{cls}), one 
finally gets
\begin{equation}
\de_{{}_M} J^{{}_M} =\frac{{g_5}^2}{32 \pi^2}  \, f(y) \, F_{\mu \nu} 
\widetilde{F}^{\mu \nu} \, , 
\qquad f(y) = \sum_m \left(\phi^{R}_m \, \phi^{R}_m \, - \phi^L_m \phi^L_m \right) \quad  .
\label{anom}
\end{equation}
It should be stressed that in deriving (\ref{anom}) one is tacitly
 supposing that 
(\ref{cls}) is still valid at the quantum level and all the quantum effects are encoded in 
Eqs. (\ref{q1}-\ref{q2}); we will show that actually this is not the case. 

The sum over the KK modes in (\ref{anom}) can be computed and it reads 
in the case ${\cal P} = \mathbb{Z}_2 
\times \mathbb{Z}_2^\prime$
for the various choices of parity for the fermion
\begin{align}
&f^{(++)}(y) = - \, f^{(--)}(y) = \frac{1}{4 R} \sum_{n=- \infty}^{+ \infty} \delta(y/R -  n \pi/2)
\quad ; \label{up}\\
&f^{(+-)}(y) = - \, f^{(-+)}(y) = \frac{1}{4 R} \sum_{n=- \infty}^{+ \infty} (-1)^n \, 
\delta(y/R -  n \pi/2) \quad .
\label{z2}
\end{align}
In particular, if the fermions have opposite parities $(+,-)$ and $(-,+)$, one recovers
Eq. (\ref{two}). For the case ${\cal P} = \mathbb{Z}_2$ one finds
\begin{equation}
f^{(+)}(y) = - \, f^{(-)}(y) = \frac{1}{2 R} \sum_{n= - \infty}^{+ \infty} \delta(y/R -  \pi n)
\quad ,
\label{z1}
\end{equation} 
which reproduces Eq. (\ref{one}). \footnote{Eqs. (\ref{up})-(\ref{z1}) still hold when a bulk mass
term $M(y)$ for the 5D fermion is present. Indeed, one can compute $f(y)$ simply by using 
the completeness of the KK modes without relying on their explicit form.}

\section{Deformed Chern-Simons counterterm}

In general, the manifestation of an  anomaly at the quantum level reflects 
the failure of removing  the 
ultraviolet divergences and -- at the same time --
preserving all the classical symmetries of the theory.

The natural question is therefore whether there exists 
a local counterterm $S_{{\rm ct}}$ 
such that the new action $S^\prime=S_{5D}+ S_{{\rm ct}}$ leads to a 
conserved 5D vector current in the orbifold theory $S^1/(\mathbb{Z}_2\times \mathbb{Z}^\prime_2)$
maintaining the symmetries of classical action.  

Consider the vacuum functional $Z[A]$
\begin{equation}
Z[A] = e^{i W[A]} = \int D[\Psi] \, e^{i \, S[A, \, \Psi]} \quad ;
\end{equation}
a would-be anomaly $G$ shows up in a non-zero variation of the 
connected generating functional $W$
under a gauge transformation $\delta A_{{}_M} = - \de_{{}_M} \lambda$ of the the vector 
potential
\begin{equation}
\delta_\lambda W[A] = i  \int d^5x \, \lambda \, G[A] \quad .
\end{equation}
As a result, one has for the divergence of the current
\begin{equation}
\de{{}_M} \langle J^{{}_M} \rangle_{\text{conn.}} = G[A] \, , \qquad 
\frac{\delta W[A]}{\delta A_M} = \langle J^{{}_M} \rangle_{\text{conn.}} \; .
\end{equation}
$G$ is defined modulo a local functional of $A$, and the presence of a true anomaly is 
related to the impossibility of finding a suitable $S_{{\rm ct}}$ such that $G$ is zero. 
The natural candidate for $S_{{\rm ct}}$ 
in our case is the following deformed Chern-Simons term
\begin{equation}
S_{{\rm CS}} = \int u(y) \, A \wedge F \wedge F = \int d^5x \, \epsilon^{{}_{MNOPQ}} A_{{}_M} \, 
F_{{}_{NO}} \, F_{{}_{PQ}} \, u(y) \quad .
\label{cs}
\end{equation}
The counterterm $S_{{\rm CS}}$ will be consistent with the orbifold projections if 
$u$ has parity $(-,-)$. 
The gauge variation of the new connected generating functional is given by
\begin{equation}
\delta_\lambda W^\prime = i  \, \int d^5x \, \lambda \left[G[A] + F_{\mu \nu} 
\widetilde{F}^{\mu \nu} \, \de_y u(y) \right] \quad ,
\end{equation}
and the condition of vanishing divergence for the current gives the following constraint on the 
function $u(y)$
\begin{equation}
\de_y u(y) + \frac{g_5^2}{32 \pi^2} \, f(y) = 0 \quad .
\label{acanc}
\end{equation}
In the case $(+,-)$ and $(+,-)$ a solution is easily found
\begin{equation}
u^{(+-)}(y) = - u^{(-+)}(y) = - \frac{g_5^2}{128 \pi^2} \, \frac{1}{2} 
\, \text{sgn}_\pi(\pi/2-y/R) \quad ;
\end{equation}
where $\text{sgn}_\pi(x)$ is the sign function periodically extended with period $\pi$.
The other cases are more subtle; if we suppose that $u(y) = u(y+ 2 \pi R)$ then Eq. (\ref{acanc})
can be solved only if
\begin{equation}
\int_0^{2 \pi R} f(y) \, dy = 0 \quad .
\end{equation}
As a result, there is no periodic $u$ which solves Eq.(\ref{acanc})
in the cases of fermions with parities $(+,+)$ $(-,-)$ 
where a chiral zero mode is present in the spectrum, the same
 result holds for   the case of a 
single $\mathbb{Z}_2$.

Since the CS term (\ref{cs}) one needs to add in order to 
preserve the gauge invariance of the theory  depends on the odd 
function $u(y)$, it becomes clear that gauge invariance can be 
maintained if one accepts the presence of odd operators which requires   
piece-wise constant (odd) functions as generalized couplings in order to 
be consistent with orbifold symmetries. 
An odd operator appears naturally when the theory is regulated in 
a gauge invariant way as a topological term in fifth component of the the 
ground-state current $\langle J^M\rangle$. The fact that an odd operator 
can have a VEV is not is not in contradiction with the invariance of the 
vacuum under the orbifold projections, indeed, due the breaking of 
translation invariance invariance one can only deduce that the VEV is an 
odd function of $y$.
As we shall demonstrate in the next section, the CS term
term can be generated naturally with the proper coefficient and sign
adopting a gauge invariant regularization.

The possibility of 
adding to the Lagrangian a CS term with an odd function $u(y)$ and a
suitable coefficient was first mentioned in 
Ref. \cite{twoz}, where 
the odd function needed in the CS term was understood
as the expectation value of a new odd field $\chi(y)$ added to the original
theory to cancel the anomaly. However our point of view is different: the theory with a single 
5D Dirac fermion is by itself non-anomalous, and there is no need to introduce an ad-hoc extra 
field. 

\section{The vacuum current}

We now derive an explicit expression for the induced vacuum current
in the orbifold theory  $S^1/(\mathbb{Z}_2\times \mathbb{Z}^\prime_2)$
for a 5D fermion with parity $(+-)$ and $(-+)$ and for simplicity with 
vanishing bulk mass. We 
consider a gauge field background which is uniform in the bulk.
The generic result can then be derived by general covariance arguments.

From the form of the
CS term, one expects that the vacuum
expectation value of the vector current acquires a non-vanishing component
along the fifth direction 
\begin{equation}
\langle J^5(x,y)\rangle=-\frac{i}{Z[A]}\,\frac{\delta Z[A]}{\delta 
A_5(x,y)}\, .
\label{vev}
\end{equation}
From this term one can  then deduce the CS term in the effective action. 
We first introduce a new fermion with  five-dimensional mass term $M(y)\overline{\Psi}\Psi$ which
at the end of the computation we decouple from the theory by taking
the mass to infinity (Pauli-Villars regularization).
Since the fermionic bilinear $\overline{\Psi}\Psi$ is odd under both the
$\mathbb{Z}_2$ and the $\mathbb{Z}^\prime_2$ parities, the function $M(y)$
has to be odd under both reflection symmetries. 
For instance, we can choose $M(y)=\kappa\, \text{Sgn}_\pi(\pi/2-y/R)\, 
, \quad \kappa > 0$, where $[\kappa]=1$. Eventually we will let $\kappa\rightarrow\infty$. 
With a mass term the equations for the KK modes read
\begin{eqnarray}
&&  \phi^{{}_R}_n \, M_n   - M(y)  \, \phi^{{}_L}_n - \, \frac{d}{dy} \phi^{{}_L}_n = 0 \quad ;
\label{k1} \\
&& \phi^{{}_L}_n \, M_n - M(y)  \, \phi^{{}_R}_n+ \, \frac{d}{dy} \phi^{{}_R}_n = 0 \quad .
\label{k2}
\end{eqnarray}
In the limit of large $\kappa$  one can infer from
Eqs. (\ref{k1}) and (\ref{k2}) that the mass eigenvalues $M_n$ satisfy the
relation
\begin{equation}
\tan\left(\frac{\pi}{2}\alpha_n R\right)=-\frac{\alpha_n}{\kappa}\, ,
\end{equation}
where 
\begin{equation}
M_n^2 = \alpha_n^2 + \kappa^2\, .
\end{equation}
The mass eigenvalues reduce to $M_n^2=
\left(\frac{2n}{R}\right)^2 +\kappa^2$ and  the corresponding 
eigenfunctions
are given by
\begin{eqnarray} 
\phi_n^{{}_L}&=&\frac{1}{\sqrt{\pi R}} \sin \left[\frac{2n}{R}y
\right] \, ,
\nonumber\\
\phi_n^R&=&\frac{1}{M_n \sqrt{ \pi R}}
\left\{M(y) \sin\left[\frac{2n}{R}y\right] +\frac{2n}{R} 
\cos\left[\frac{2n}{R}y\right]\right\}\, .
\label{v}
\end{eqnarray} 
The crucial point is that the eigenfunction $\phi_n^R$ picks up a piece
which depends upon the odd mass term $M(y)$.

Our goal is
to show that the decoupling procedure leaves a finite CS term in the
regularized action $S_{{\rm reg}}=S_{5D}[A]-S_{5D}[A,M]$ when
the five-dimensional mass is taken to infinity.

Consider the vacuum expectation value of $J^5$ in the external gauge field
\begin{equation}
\langle \, J^5(x,y) \, \rangle = -i \sum_{mn} \phi_m^{{}_L}(y) \phi_n^{{}_R}(y) \; 
\langle \, j_{5\, mn}(x) \, \rangle\quad .
\end{equation}
The relevant contribution to the expectation value of $j_5$ is the one-loop diagram 
in Fig. 1 where the fermions $\psi_{n{}_{L/R}}$ run in the loop. Selecting the part which will
give the finite piece in the large $\kappa$ limit, one has 
\begin{equation}
\begin{split}
\langle \, j_{5\, mn}(x) \, \rangle &= - \delta_{mn} \frac{g_5^2}{2} \int \frac{d^4Q_1}{(2 \pi)^4} 
\frac{d^4Q_2}{(2 \pi)^4} \frac{d^4q}{(2 \pi)^4} \, e^{-i x \cdot(Q_1+Q_2)} \, A_\mu(x) \, 
A_\nu(x)\\  
&\frac{8 i \, \epsilon^{\mu \alpha \nu \beta} \,  Q_{1 \alpha} \, Q_{2 \beta} \, M_n }
{\left[(Q_1+q)^2-M_n^2 + i \epsilon \right] \left(q^2 -M_n^2 + i \epsilon \right) \left[(q-Q_2)^2-M_n^2 + i \epsilon \right]}
\end{split}
\label{mmm}
\end{equation}

\begin{figure}[t]  
\begin{center}  
\begin{picture}(200,200)(0,0)  
\SetOffset(-30,0)  
\Text(25,170)[c]{${j_{{}_5}}_{mn}(x)$}
\Vertex(20,150){3}
\ArrowLine(20,150)(-30,100)
\Vertex(-30,100){3}
\Photon(-30,100)(-50,80){3}{4}
\ArrowLine(-30,100)(70,100)
\Vertex(70,100){3}
\ArrowLine(70,100)(20,150)
\Vertex(20,150){3}  
\Photon(70,100)(90,80){3}{4}  
\Text(100,70)[c]{$j^{{}_{L/R}}_{\mu \, op}(x_1)$}  
\Text(-55,70)[c]{$j^{{}_{L/R}}_{\nu \, rs}(x_2)$}    
\SetOffset(190,0)  
\Text(25,170)[c]{${j_{{}_5}}_{mn}(x)$}
\Vertex(20,150){3}
\ArrowLine(20,150)(-30,100)
\Vertex(-30,100){3}
\Photon(-30,100)(-50,80){3}{4}
\ArrowLine(-30,100)(70,100)
\Vertex(70,100){3}
\ArrowLine(70,100)(20,150)
\Vertex(20,150){3}  
\Photon(70,100)(90,80){3}{4}  
\Text(100,70)[c]{$j^{{}_{L/R}}_{\nu \, rs}(x_2)$}
\Text(-55,70)[c]{$j^{{}_{L/R}}_{\mu \, op}(x_1)$}    
\end{picture}  
\end{center}  
\vskip -1truecm  
\caption{Triangle diagrams contributing to the vacuum expectation value of $J^5$}  
\label{model}  
\end{figure}
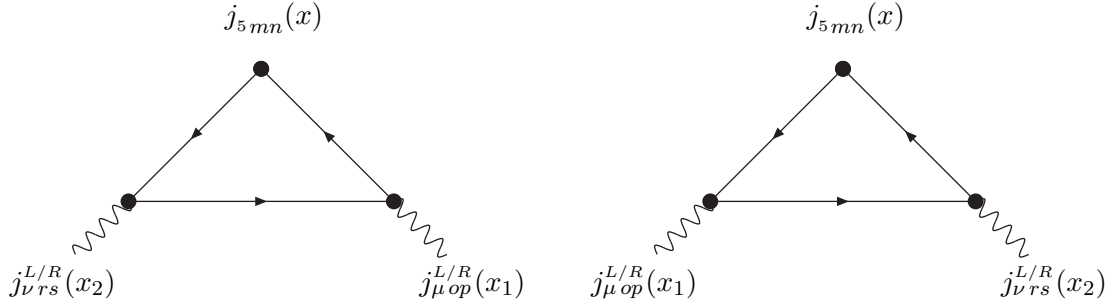

The fact that the eigenfunction $\phi_n^{{}_R}$ has a piece proportional 
to $M(y)$ gives rise to a finite non-vanishing contribution 
left-over in the large
$\kappa$ limit. Indeed, in this limit, the external momenta in the denominator of (\ref{mmm}) 
can be set to zero, and we get 
\begin{equation}
\langle \, J^5(x,y) \, \rangle = - \frac{g_5^2}{32 \pi^2} \, F_{\mu \nu} \widetilde{F}^{\mu \nu}
\, \sum_n \frac{1}{M_n} \phi_n^{{}_L}(y) \phi_n^{{}_R}(y) \quad .
\end{equation}
Using the approximate form (\ref{v}) for the modes and the relevant part in the large $\kappa$ 
limit, we have
\begin{equation}
\langle \, J^5(x,y) \, \rangle = - \frac{g_5^2}{32 \pi^2} \, F_{\mu \nu} \widetilde{F}^{\mu \nu}
 \, M(y) \, \frac{ \text{Tanh}(\kappa \pi R/2)}{8 \kappa} \quad .
\end{equation}  
The ratio $ M(y)/\kappa$ does not depend upon $\kappa$ 
and changes sign around the fixed points. We can safely take the
limit $\kappa\rightarrow\infty$ and find
\begin{equation}
\langle \, J^5(x,y) \, \rangle = - \frac{g_5^2}{32 \pi^2} \, \text{Sgn}_\pi \left(\pi/2 - y/R 
\right) \, \frac{1}{8} \, F_{\mu \nu} \widetilde{F}^{\mu \nu} \quad .
\end{equation}
Thus, the physical ground-state gets a VEV  and we
discover that  a finite CS counterterm is produced    
\begin{equation}
{\cal L}_{{\rm CS}}= -
\frac{g_5^2}{128 \pi^2}\,\frac{1}{2}\,\text{Sgn}_\pi(\pi/2-y/R)\, A_5(x,y)\,
\,F_{\mu\nu}(x,y)\widetilde{F}^{\mu\nu}
(x,y)\, 
\end{equation}
which is the precisely the form  needed in the expression (\ref{cs}) to
cancel the anomaly. As a result, the cancellation of the divergence in 
Eq. (\ref{anom}) may be thought as coming from 
$\langle \partial_y J^5\rangle$, which can be non-vanishing due the lack
of translation invariance in the fifth dimension.
Consequently, $\langle \partial_y J^5\rangle$ gets an extra 
contribution because of the nontrivial shape of the function 
$\partial_y M(y)$. 

Let us close with some comments about theories with a single  
$\mathbb{Z}_2$. The reader might expect that the relation between 
spontaneous breaking of the reflection symmetry and gauge
symmetry leading to the cancellation of the anomaly 
described in this paper might work in that case too. However, this is 
not the case. The basic difference is that in 5D theories compactified 
on $S^1/\mathbb{Z}_2$ the odd function
that one should introduce in the CS term to reproduce the singular 
structure in Eq. (\ref{one}) is monotone in the bulk space with jumps 
every $\pi R$. The problem is that such a function  is not periodic. 
Notice also that, with a single orbifold projection, when
we take the limit $\kappa \to \infty$, the fermion does not decouple due 
to the chiral 
zero mode whose bulk wave-function becomes peaked 
at  one of the fixed points.

\section{Conclusions}
In this paper we have studied the issue of gauge
invariance
in models with five-dimensions 
compactified on an orbifold $S^1/(\mathbb{Z}_2\times \mathbb{Z}^\prime_2)$. If, upon
reduction to four-dimensions, the theory contains a tower of KK chiral fermions 
with opposite parities under the two $\mathbb{Z}_2$'s, the theory looks
anomalous at the quantum level. This would be anomaly can be canceled by a suitable
topological CS term which is made even under the two reflection 
symmetries by the presence of an odd  piece-wise constant function which 
is allowed by the orbifold symmetries.
We have shown that such a CS term is generated if the 
gauge theory is regularized in a gauge invariant way 
This occurs because 
the  fermionic current acquires a vacuum expectation value induced 
by the classical background gauge field. This phenomenon is 
analogous to what happens in theories with an odd number of non-compact 
dimensions where the parity-violating part of the vacuum current induced 
by a classical background gauge field implies the presence in the
effective action of a CS topological invariant which is odd under parity
transformations.

\vskip 0.2cm
\centerline{\bf Acknowledgments}
\vskip 0.2cm
We thank F. Zwirner for useful discussions and especially R. Barbieri, R.
Contino and R. Rattazzi for letting us know about
their work in progress on similar issues.
Work supported in part by the RTN European Program HPRN-CT-2000-00148

\end{document}